\documentclass[aps,pra,reprint,superscriptaddress]{revtex4-1}

\usepackage{lineno}

\usepackage{hyperref}
\usepackage{graphicx}  
\usepackage{dcolumn}   
\usepackage{bm}        
\usepackage{amssymb}   
\usepackage{epstopdf}
\usepackage{amsmath}
\usepackage{xspace}
\usepackage{verbatim}
\usepackage{color}

\usepackage{soul}

\newcommand{\bra}[1]{\left\langle #1 \right|}
\newcommand{\ket}[1]{\left| #1 \right\rangle}

\hypersetup{
    colorlinks=true,linkcolor=blue,citecolor=blue,
    filecolor=blue,urlcolor=blue,breaklinks=true
}

\begin{document}
\title{Weak value amplification of photon number operators in the optomechanical interaction}

\author{Sergio Carrasco}
\email{sjcarras@uc.cl}
\affiliation{Instituto de F\'{i}sica, Pontificia Universidad Cat\'{o}lica de Chile,
Casilla 306, Santiago, Chile}

\author{Miguel Orszag}
\affiliation{Instituto de F\'{i}sica, Pontificia Universidad Cat\'{o}lica de Chile,
Casilla 306, Santiago, Chile}
\affiliation{Universidad Mayor, Avda. Alonso de C\'ordova 5495, Las Condes, Santiago, Chile}

\date{\today}

\begin{abstract}
An experimental proposal is presented in which dark port post-selection together with weak measurements are used to enlarge the radiation pressure effect of a single photon on a mechanical oscillator placed in the middle of a Fabry-P\'erot cavity and initialized  in the ground state.  By preparing and post-selecting the photon (the system) in two quasi orthogonal  states, the weak value of the radiation force operator can lie outside the eigenvalue spectrum, producing a large shift on the wave function of the mechanical oscillator (the measuring device) in the position representation.  Consequently, the effect of a single photon on the average position of the oscillator in its final state can be amplified as compared to the effect caused by a photon without post-selection, i.e. only pre-selected. The strong  measurement scenario is also analyzed. In this case, a higher amplification effect is achieved and the mean position of the oscillator reaches the level of the zero-point fluctuation, but the back-action on the system is increased and the post-selection probabilities are smaller.
\end{abstract}

\maketitle

\section{Introduction}

Weak measurements \cite{Svensson2013,TamirCohen2013,Kofman2012,Dressel2014,QuantumParadoxes} correspond to  standard von Neumann measurements in the regime in which the interaction strength between the system and the apparatus is weak enough so that the back-action on the system can be neglected, while the measurement still provides a small amount of information. This occurs due to the fact that, when the interaction is weak, perturbations of the initial system state depend quadratically on the interaction strength, whereas the information depends linearly \cite{Tollaksen2010}.

The ensemble average of weak measurements, given an initial system state, is just the expectation value of the observable being measured, i.e. the same result as for projective measurements. When an initial and a final state are given, the ensemble average of a weak measurement  is called a \textit{weak value} \cite{AAV1988}. In this case, ``neglecting the back-action'', which is the defining feature of a weak measurement, means that both, the initial and the final states of the system, affect the measuring device in the same way, i.e. both enter linearly in the definition of the weak value.  Moreover, when these two states are nearly orthogonal, the weak value can lie outside  the eigenvalue spectrum of the measured variable at the expense of reducing the post-selection probability.  In this case, since the weak value is larger than any of the eigenvalues, there is an amplification effect and the wave function of the measuring device, in some proper representation, is displaced by a large amount, proportional to the weak value. This technique has been called weak value amplification (WVA), while weak values lying outside the spectrum are often referred to as \textit{anomalous}  or \textit{strange} weak values \cite{Sokolovski2016,Hosoya2010}. This term might also be used for negative weak values of number operators or for complex weak values \cite{Jozsa2007}.  When the measurement is strong (and post-selection is performed), the amplification may also occur but in this case the back-action effect on the system is not negligible, and  thus the pre and post-selected states do not affect the measuring device linearly. 
 
Weak measurements and weak values have been employed to study fundamental issues in quantum mechanics, such as wave-particle duality \cite{Aharonov2017},  non locality \cite{WeakValuesNonLocality2016} and contextuality \cite{AharonovNewIO}.   Weak values have also been used to analize the quantum shell game \cite{Tavon2007,Resch2004}, Cheshire's cats \cite{Aharonov2013,Denkmayr2014} or Hardy's paradox \cite{Hardy1992,HardyAharonov2002,HardyExp2009}. On the other hand, since their formulation weak values have overcome different controversies \cite{WVControversy2017}. In \cite{FerrieCombes2014} it is argued that weak value amplification could be explained classically. However, it turns out that it is a truly quantum effect \cite{Mundarain2016}.  

From a technological point of view,  WVA has proven to be useful for the estimation of small parameters, e.g. angular displacements of a tilted mirror $\sim$ prad \cite{Dixon2009} or small changes in frequency \cite{Starling2010}.  It has been employed to enlarge the transverse separation of an optical beam due to birefringence \cite{Ritchie1991,Hall2008}, longitudinal phase shifts \cite{Brunner2010} or angular rotations of a classical beam \cite{Loaiza2014}, among others. WVA has also allowed a direct observation of the quantum wave function \cite{Lundeen2011} and to observe the average trajectories in a two slit interferometer \cite{Kocsis2011}. In \cite{Feizpour2011,Feizpour2015,Hallaji2017} Feizpour \textit{et al.} showed that the non linear cross-Kerr effect of a single photon on a classical beam can be amplified and observed using weak measurements. In \cite{Simon2011} an implementation of WVA with photons and atomic ensembles was proposed, as a method for measuring the small coupling constant between light and the ensemble of atoms.  Recently, weak measurements have been theoretically proposed for quantum control applications \cite{Coto2017} and for the preparation of macroscopic non classical states \cite{Montenegro2017}. 

In this paper, we show that weak measurements followed by post-selection can be used to enlarge the radiation pressure effect of a single photon on a mechanical resonator placed in the middle of an optical cavity, as compared to the scenario with no post-selection.  In our setup, the radiation force is proportional to the difference of photons between both sides of the cavity and hence the amplification is associated to weak values of the difference of photon number operators. We show that an anomalous weak value will produce a large displacement of the oscillator's position and compare the WVA to the scenario in which the level of back-action is increased, i.e. when the measurement gets stronger.  From a practical side, the proposed scheme might be used for precision measurements of the vacuum optomechanical strength, when the mechanical oscillator is operated at the quantum level, i.e. when it is in the ground state or in some linear combination of low number states. 

The structure of this article is as follows. In section \ref{sec:WV} we present for completeness a quick review of weak measurements and weak values. In section \ref{sec:Model} we describe the experimental proposal, present the hamiltonian model and the time evolution operator that allows to perform deterministic evolution of the system. In section \ref{sec:WVA} we show the WVA effect and compare it to the amplification factor obtained when the level of back-action is increased. Finally, in section \ref{sec:FinalRemarks}, the results are summarized and commented. 

\section{Weak Measurements and Weak Values}\label{sec:WV}

According to \cite{vonNeumann} the measurement process of a physical variable $\hat{A}$ can be described by a hamiltonian of the form $\hat{H}=\hat{H}_0+g(t)\hat{A}\hat{p}$, where $\hat{H}_0$ is the unperturbed hamiltonian of the system and the apparatus, while the second term describes the interaction between them during the measurement.  The coupling $g(t)$ is an impulsive function that depends on time and that is switched on during the interaction and then turned off, while  $\hat{p}$ is a physical variable of the meter, conjugate to $\hat{q}$.  The system has initially no correlation with the apparatus and consequently the initial state is described by the product $\ket{\Phi(0)}=\ket{i}\ket{\psi}_{m}$, where $\ket{i}$ is the initial state of the system and $\ket{\psi}_{m}$ the initial state of the meter.  When the interaction term commutes with $\hat{H}_0$, the time evolution operator, in an interaction picture with respect to $\hat{H}_0$, is  $\hat{U}=\exp{(-i\frac{g}{\hbar}\hat{A}\hat{p})}$, where $g$  is just the integration of $g(t)$ over its compact support.  Assuming that the state $\ket{f}$ is post-selected, then the unnormalized state of the apparatus may be expressed as
\begin{eqnarray}\nonumber
\ket{\phi}_{m}&=&\bra{f}\hat{U}\ket{i}\ket{\psi}_{m}\\ \nonumber
&\approx&\bra{f}[1-i(g/\hbar)\hat{A}\hat{p}]\ket{i}\ket{\psi}_{m}\\ \nonumber
&=&\langle f|i\rangle[1-i(g/\hbar)A_{w}\hat{p} ] \ket{\psi}_{m}\\ \label{DerivationofWV}
&\approx&\langle f|i\rangle\exp{[-i(g/\hbar)A_{w}\hat{p}]}\ket{\psi}_{m}.
\end{eqnarray}

The conditions under which the approximations above hold are analyzed in \cite{Duck1989}. The complex number $A_{w}$ is defined as
\begin{eqnarray}\label{eq:TheWV}
A_{w}=\frac{  \bra{f}\hat{A}\ket{i}    }  {\langle f | i \rangle  },
\end{eqnarray}
and is called the \textit{weak value} of the operator $\hat{A}$. When the weak value is real the operator $\exp{[-i(g/\hbar)A_{w}\hat{p}]}$ produces a translation of the wave function in the $q$-representation by an amount of $gA_{w}$.  Consequently, the wave function of the meter preserves its shape and is only translated. In some situations, the coupling constant $g$ is a small unknown parameter to be estimated. In this work, this parameter will be proportional to the vacuum optomechanical strength. 

\section{THE MODEL}\label{sec:Model}

\subsection{Description of the Experiment}\label{subsection-a}

\begin{figure}
 \centering \includegraphics[width=\linewidth]{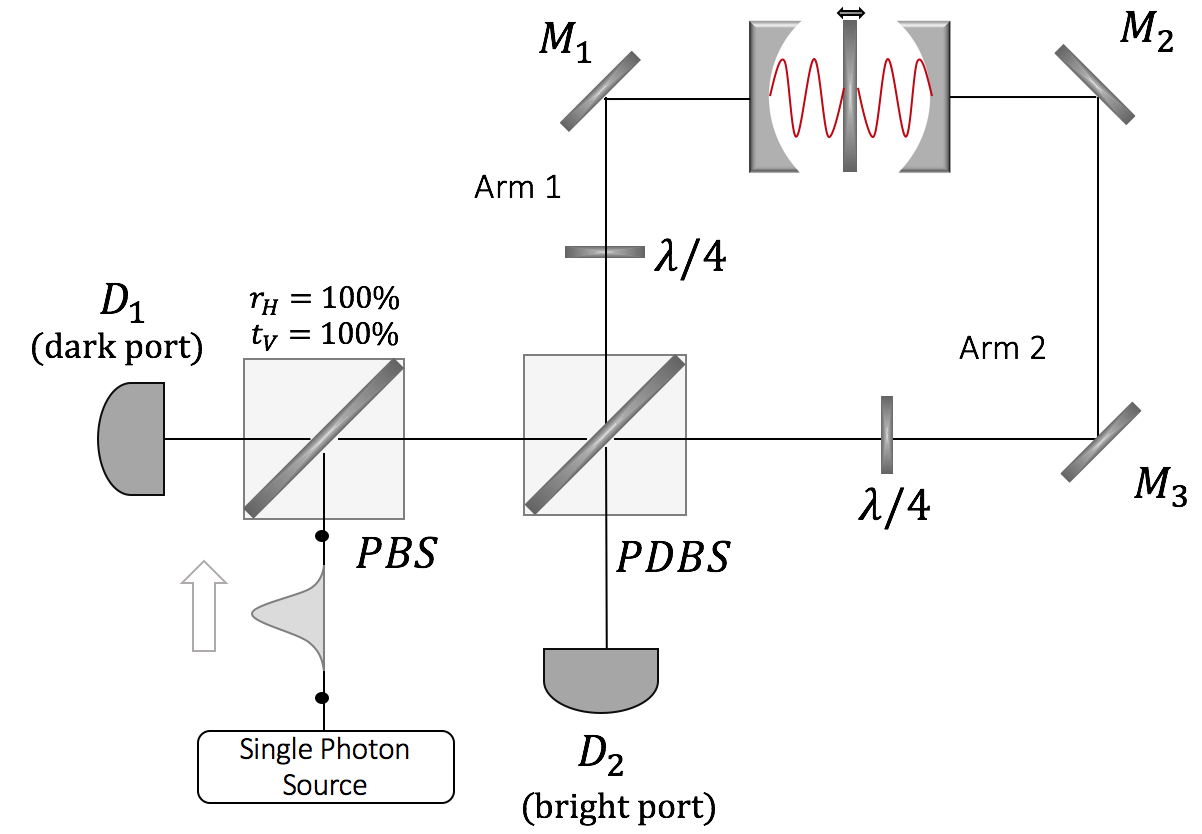}
 \caption{Optomechanical system (OM) in an interferometer. The OM system is a cavity with an oscillator in the middle, which is assumed to be perfectly reflecting. The PBDS is balanced for horizontal polarization and has an imbalance $\delta$ for vertical polarization. Photon counters $D_1$ and $D_2$ are used as detectors at the output ports of the interferometer.} \label{fig:fig1}
\end{figure}

The proposed setup consists of an optomechanical system mounted inside an interferometer (figure \ref{fig:fig1}). The optomechanical system is an optical cavity with a high-Q mechanical oscillator in the middle. The oscillator is considered to be a perfectly reflecting mirror and therefore the cavity modes of each side do not interact directly (no tunneling between both cavity modes is allowed). Since the mirror is put exactly in the middle, both optical modes have the same frequency $\omega_0=n_0c\pi/L$, where $c$ is the speed of light, $L$ is the effective length of each side of the cavity and $n_0$ is the integer mode number.    

The experimental proposal works as follows. The oscillator is prepared in the ground state \cite{Bhaktavatsala2016,Connell2010,Teufel2011,Peterson2016}. On the other hand, one single photon, prepared with horizontal polarization, should be injected into the system through the input port. After being reflected by a polarizing beam splitter (PBS), with 100$\%$ reflectivity for horizontal polarization and $100\%$ transmissivity for the vertical component, the photon enters the interferometer through a polarization dependent beam splitter (PDBS), which is balanced for horizontal polarization. The photon travels then along the arms of the interferometer and interacts with the cavity and the oscillator, returning back to the PDBS in an entangled state with the vibrating mirror.  The photon comes back with vertical polarization due to the action of the $\lambda/4$ plates mounted on each arm of the interferometer.  For vertical polarization, the PBDS is unbalanced. This imbalance is accounted by the parameter $\delta\equiv(t-r)/\sqrt{2}$, where $t$ and $r$ are the  transmission and reflection coefficients for vertical polarization, respectively. The coefficients are taken to be real and positive, and satisfy $r^2+t^2=1$.

After exiting, the photon can be detected in $D_1$ or  $D_2$. If $\delta=0$, then no light would arrive at  $D_1$ and therefore this port is called the \textit{dark port}.  Successful post-selection occurs when a photon is detected in  this port. The cases when post-selection fails should be disregarded and, in this situation, the quantum state of the oscillator should be reinitialized in the ground state, but when post-selection is successful the position of the oscillator should be observed \cite{Clerk2008,Vanner2011}.  Under certain conditions, described along this work, the average value of the position will correspond to the \textit{weak value} of the radiation force operator.

The experiment is designed for single photons with frequency $\omega_0$. The photons are considered to be nearly monochromatic, i.e. with an spectral bandwidth $\epsilon$ much smaller than the cavity decay rate $\gamma$ (the time duration of the photon is greater than the cavity mean storage time). 

\subsection{Hamiltonian Model}\label{subsection-b}

The hamiltonian for the whole system can be written as a sum of two contributions, namely, 
\begin{eqnarray}\label{eq:TotalHamiltonian}
\hat{H}=\hat{H}_{OM}+\hat{H}_{\rm{ext}}.\end{eqnarray}
The first term is the optomechanical hamiltonian for the \textit{closed} system comprised by the cavity and the oscillator. It can be written as
\begin{eqnarray}\label{eq:HOM}
\nonumber\hat{H}_{OM}&=&\hbar\omega_0(\hat{a}^{\dagger}_1\hat{a}_1+\hat{a}^{\dagger}_2\hat{a}_2)+\hbar\omega_{m}\hat{c}^{\dagger}\hat{c}\\&&-\hbar g_0(\hat{a}_1^{\dagger}\hat{a}_1-\hat{a}_2^{\dagger}\hat{a}_2)(\hat{c}^{\dagger}+\hat{c}).
\end{eqnarray}
In this expression, the first and second terms correspond to the free energy of the cavity and the oscillator, respectively. The third term represents the optomechanical interaction via radiation pressure. The operators $\hat{a}_1$ ($\hat{a}^{\dagger}_1$) and $\hat{a}_2$  ($\hat{a}^{\dagger}_2$) are boson annihilation (creation) operators for the cavity modes of the left and right sides, respectively. Similarly, $\hat{c}$ ($\hat{c}^{\dagger}$) is the mechanical annihilation (creation) boson operator and $\omega_{m}$ is the mode frequency of the oscillator. The parameter $g_0$ corresponds to the vacuum optomechanical coupling strength, which typically is a small parameter, i.e. $g_0<1$, that quantifies the interaction strength between one photon and one phonon inside the cavity. This parameter can be written as $x_0G$, where $G=\omega_0/L$ is the frequency shift per displacement and $x_0$ is the mechanical zero-point fluctuation.  In the third term, a minus sign appears between the number operators because the movement of the mirror in one direction shortens the effective length of one side of the cavity while enlarges the other. The radiation force is given by $\hat{F}_{\rm{rad}}=\hbar G(\hat{a}_1^{\dagger}\hat{a}_1-\hat{a}_2^{\dagger}\hat{a}_2)$ \cite{Optomechanics} and consequently this term can be compactly expressed as $-\hat{F}_{\rm{rad}}\hat{q}$, where $\hat{q}$ is the position of the oscillator. 

\begin{figure}
 \centering \includegraphics[width=\linewidth]{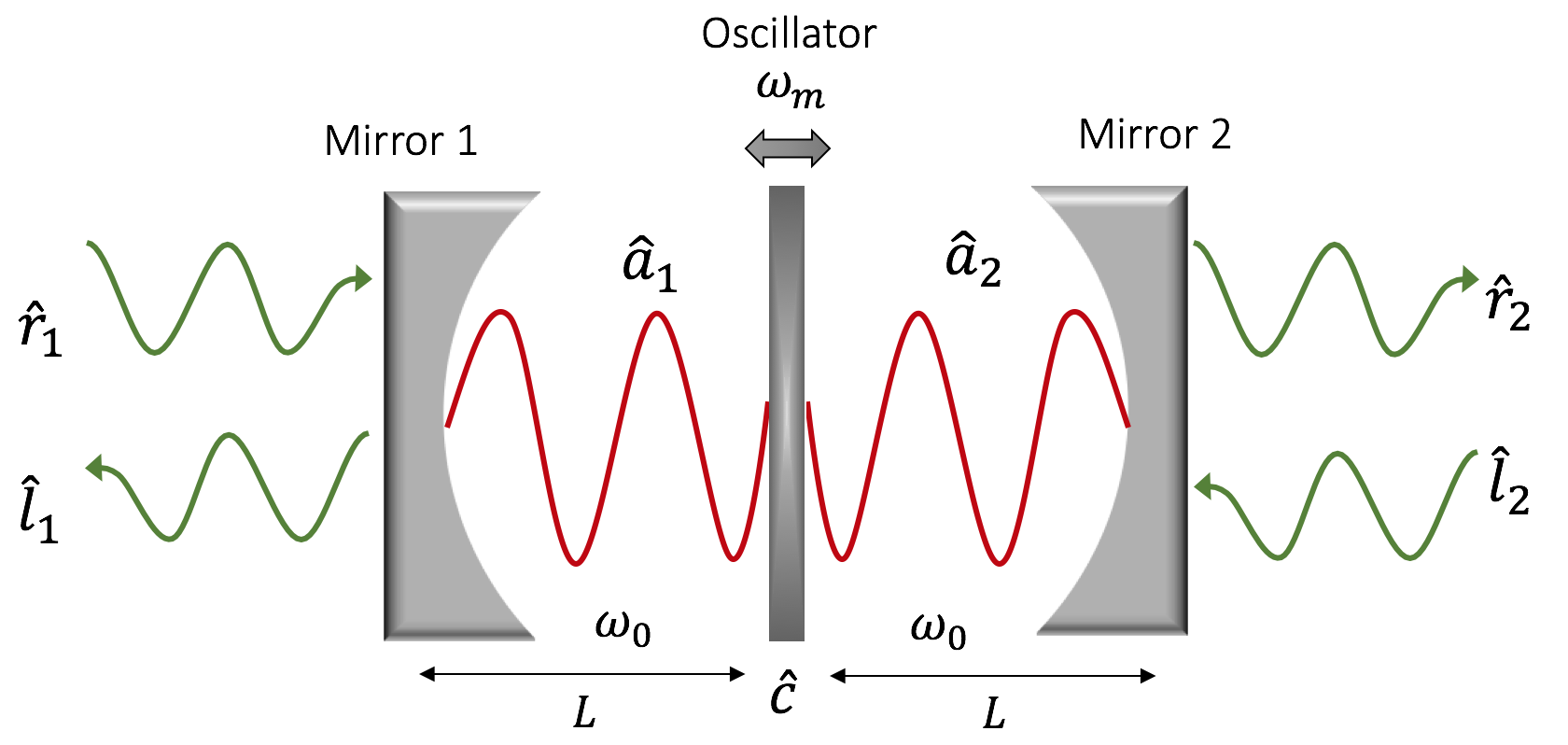}
 \caption{A cavity with a moving mirror in the middle. The oscillator does not allow transmission of light from one side to the other. The optical modes of each side interact with the oscillator through radiation pressure. In turn, the cavity interacts with the external field because the mirrors are semi transparent.} \label{fig:fig2}
\end{figure}

The second term in (\ref{eq:TotalHamiltonian}) describes the energy of the external field and its interaction with the cavity modes.  The external field is a continuum of modes, but, since the injected photon has a well defined frequency and thus a  long time duration (which would be infinite in an idealized theoretical scenario), this term will be described according to the model
\begin{eqnarray}\nonumber
\hat{H}_{\rm{ext}}=
\hbar\omega_0\sum_{i=1}^{2}(\hat{r}^{\dagger}_{i}\hat{r}_{i}+\hat{l}^{\dagger}_{i}\hat{l}_{i})+
\hbar \xi\sum_{i=1}^2\Big[\hat{a}_i (\hat{r}^{\dagger}_i+\hat{l}^{\dagger}_i)+
\text{h.c.}\Big].\\ \label{eq:Hext}
\end{eqnarray}
The first term is the free energy of the external field and the second describes its interaction with the cavity. The operators $\hat{r}_{1}$ ($\hat{r}^{\dagger}_{1}$) and $\hat{r}_{2}$ ($\hat{r}^{\dagger}_{2}$) are annihilation (creation) operators for travelling modes propagating to the \textit{right},  through the left and right arms, respectively. Analogously, $\hat{l}_{1}$ ($\hat{l}^{\dagger}_{1}$) and $\hat{l}_{2}$ ($\hat{l}^{\dagger}_{2}$) are annihilation (creation) boson operators for modes propagating to the \textit{left}, in the left and right arms, respectively (see figure \ref{fig:fig2}).  The parameter $\xi$ depends on the photon-hopping interaction strength between the cavity and the external field (which accounts for transmission losses through the cavity mirrors) and on the spectral bandwidth of the external field.  Optical damping through other channels and mechanical damping are assumed to occur much slower and, consequently, will be neglected from the model. 

Let us define for each arm of the interferometer, i.e. for $i=1,2$, the standing-wave mode operators
\begin{eqnarray}\label{eq:SWmodes}
\hat{b}_{i}=\frac{\hat{r}_{i}+\hat{l}_{i}}{\sqrt{2}}, \quad 
\hat{d}_{i}=\frac{\hat{r}_{i}-\hat{l}_{i}}{\sqrt{2}}.
\end{eqnarray}

In an interaction picture with respect to the free energy of the cavity and the external field, the hamiltonian becomes 
\begin{eqnarray}
\nonumber\hat{H}_{\rm{I}}=\hbar \sqrt{2}\xi\sum_{i=1}^2(\hat{a}_i^{\dagger}\hat{b}_{i}+\text{h.c.})+\hbar\omega_{m}\hat{c}^{\dagger}\hat{c}\\
\label{eq:HI}
-\hbar g_0(\hat{a}_1^{\dagger}\hat{a}_1-\hat{a}_2^{\dagger}\hat{a}_2)(\hat{c}^{\dagger}+\hat{c}).
\end{eqnarray}
Notice that the cavity interacts only with even-parity modes, with a coupling constant strengthened by a factor of $\sqrt{2}$. For simplicity, this factor will be absorbed into $\xi$. Odd-parity modes evolve freely. 

When $g_0\ll\xi$, the energy inside the cavity will depend mostly on the interaction with the external field rather than the interaction with the mirror (since the optomechanical interaction is much weaker than the interaction with the external field). In this regime, the energy inside the cavity (and thus the radiation force on the oscillator) will change in a time scale given by $\xi^{-1}$.   In addition, when $\omega_m\ll\xi$, the radiation force will vary much faster than the mechanical period and the effective force will correspond to an average. Also, typically $g_0\ll\omega_m$. Consequently, in the regime defined by $g_0\ll\omega_m\ll\xi$, the hamiltonian (\ref{eq:HI}) becomes approximately 
\begin{eqnarray}\label{eq:HI2}
\hat{H}_{\rm{I}}=2\xi\hat{J}_{x}+\hbar\omega_{m}\hat{c}^{\dagger}\hat{c}-\frac{\hbar g_0}{2}\hat{N}(\hat{c}^{\dagger}+\hat{c}).
\end{eqnarray}
In this regime, the radiation force is proportional to $\hat{N}$, defined as the difference of interacting photons between both sides, namely,
\begin{eqnarray}\label{eq:Force}
\hat{N}=(\hat{a}^{\dagger}_1\hat{a}_1+\hat{b}^{\dagger}_1\hat{b}_1)-(\hat{a}^{\dagger}_2\hat{a}_2+\hat{b}^{\dagger}_2\hat{b}_2),
\end{eqnarray}  
while the first term is simply $\hat{J}_x=(\hbar/2)\sum_{i=1}^2(\hat{a}_i^{\dagger}\hat{b}_{i}+\hat{a}_i\hat{b}_{i}^{\dagger})$. A  proof of this approximation is given in appendix \ref{app:A}. Hamiltonian (\ref{eq:HI2}) preserves the total number of photons and the radiation force is a constant of motion, i.e. $[\hat{N},\hat{H}_I]=0$, unlike in the original hamiltonian (\ref{eq:TotalHamiltonian}). In addition, since $[\hat{J}_x,\hat{N}]=0$, the corresponding time evolution operator can be easily disentangled. After some calculations (see appendix \ref{app:B} for the details), the time evolution operator can be expressed as  
\begin{eqnarray}
\nonumber\hat{U}_I(\tau)&=&\exp{[i\phi(\tau)\hat{N}^2]}
\exp{\{\hat{N}[\varphi(\tau)\hat{c}^{\dagger}-\varphi^*(\tau)\hat{c}]\}}
\\ \label{eq:TimeEvolutionOperator}
&&\times\exp{[-i(2\xi \tau/\hbar)\hat{J}_{x}]}\exp{(-i\omega_m\tau\hat{c}^{\dagger}\hat{c})},
\end{eqnarray}
\noindent where $\varphi(\tau)=(g_0/2\omega_m)(1-e^{-i\omega_m\tau})$, $\phi(\tau)=(g_0/2\omega_m)^2(1-\sin\omega_m\tau)$ and $\tau$ is the time of interaction. The first term contains a non linear Kerr phase. In the subspace of single photons $\hat{N}^2=1$ and the Kerr term will merely add a global phase factor, which has no effect. Thus, for our setup, the optomechanical interaction is \textit{linear} in the number of photons $\hat{N}$. The second term  entangles the photon to the oscillator and will be denoted by $\hat{U}_{OM}(\tau)$.  The third operator describes the photon exchange process between the external (even) modes and the cavity modes, and will be called $\hat{U}_{\rm{ex}}(\tau)$. The fourth term is just the free evolution of the mechanical oscillator, and will be denoted by  $\hat{U}_{m}(\tau)$.  Therefore, the time evolution operator can be expressed compactly as $\hat{U}_{\rm{I}}(\tau)=\hat{U}_{\rm{OM}}(\tau)\hat{U}_{\rm{ex}}(\tau)\hat{U}_{m}(\tau)$. 

\subsection{Single Photon States}\label{subsec:c}

For the cavity subspace the single-photon states are defined as 
\begin{eqnarray}\label{eq:SinglePhotonStatesCavity}
\hat{a}^{\dagger}_1\hat{a}_1\ket{1,0}_{c}=\ket{1,0}_{c}, \quad
\hat{a}^{\dagger}_2\hat{a}_2\ket{0,1}_{c}=\ket{0,1}_{c}.
\end{eqnarray}
The two-mode vacuum state will be denoted by $\ket{\emptyset}_{c}$. For the external field, the single-photon states are defined according to
\begin{eqnarray}\nonumber
\hat{r}_{1}^{\dagger}\hat{r}_{1}\ket{1,0,0,0}_{T}=\ket{1,0,0,0}_{T}, \\
\nonumber
\hat{l}_{2}^{\dagger}\hat{l}_{2}\ket{0,1,0,0}_{T}=\ket{0,1,0,0}_{T}, \\ 
\nonumber
\hat{l}_{1}^{\dagger}\hat{l}_{1}\ket{0,0,1,0}_{T}=\ket{0,0,1,0}_{T}, \\
\label{eq:SinglePhotonStates}
\hat{r}_{2}^{\dagger}\hat{r}_{2}\ket{0,0,0,1}_{T}=\ket{0,0,0,1}_{T}.
\end{eqnarray}
This set of eigenvectors defines a travelling wave basis that allows to express single-photon states of the external field. The four-mode vacuum state will be denoted by $\ket{\emptyset}_{T}$. An analogous standing-wave basis can be constructed using the second set of mode operators (\ref{eq:SWmodes}).

\section{Weak Value Amplificacion}\label{sec:WVA}

Using the notation introduced in \ref{subsec:c}, the initial state of the electromagnetic field (the system) can be written as $\ket{\psi_i}=(1/\sqrt{2})(\ket{1,0,0,0}_{T}+\ket{0,1,0,0}_{T})\ket{\emptyset}_c$, while the oscillator (the meter) starts in the ground state $\ket{0}_m$. The evolved state is obtained by applying (\ref{eq:TimeEvolutionOperator}) to the product state $\ket{\psi_i}\ket{0}_m$.  By defining the time-dependent state $\ket{i}=\hat{U}_{ex}(\tau)\ket{\psi_i}$, which simply represents the evolution of the initial optical state given by the interaction between the cavity and the external field, the state of the system and the meter after the interaction can be expressed as $\ket{\Phi(\tau)}=\hat{U}_{OM}(\tau)\ket{i}\ket{0}_m$, namely,
\begin{eqnarray}\nonumber
\ket{\Phi(\tau)}=\ket{1,0,0,0}_T\ket{\emptyset}_c\Bigg[\frac{\ket{0}_m+\cos(\xi\tau)\ket{\varphi(\tau)}_m}{2\sqrt{2}}\Bigg]\\ \nonumber
+\ket{0,1,0,0}_T\ket{\emptyset}_c\Bigg[\frac{\ket{0}_m+\cos(\xi\tau)\ket{-\varphi(\tau)}_m}{2\sqrt{2}}\Bigg]\\
\nonumber
-\ket{0,0,1,0}_T\ket{\emptyset}_c\Bigg[\frac{\ket{0}_m-\cos(\xi\tau)\ket{\varphi(\tau)}_m}{2\sqrt{2}}\Bigg]\\
\nonumber
-\ket{0,0,0,1}_T\ket{\emptyset}_c\Bigg[\frac{\ket{0}_m-\cos(\xi\tau)\ket{-\varphi(\tau)}_m}{2\sqrt{2}}\Bigg]\\ \label{eq:Evolution}
-i\sin(\xi\tau)\ket{\emptyset}_T\Bigg[\frac{\ket{1,0}_c\ket{\varphi(\tau)}_m+\ket{0,1}_c\ket{-\varphi(\tau)}_m}{2}\Bigg].
\end{eqnarray}

The states $\ket{\pm\varphi(\tau)}_m$ are mechanical coherent states.  According to state (\ref{eq:Evolution}), during the evolution the photon is brought into a superposition state, simultaneously travelling into the cavity (first two terms), propagating away from it (third and fourth lines) and being inside the cavity (last term). Notice that the travelling modes produce a superposition of mechanical states, between the ground state and a coherent state. This occurs because, using equation (\ref{eq:SWmodes}), each travelling mode can be decomposed into an even-parity mode (that displaces the oscillator) and an odd-parity mode (that does not interact with the cavity). 

When the photon is detected in the dark port, the state $\ket{f}=(r\ket{0,0,1,0}_{T}-t\ket{0,0,0,1}_{T})\ket{\emptyset}_c$ is post-selected. If we assume that the interaction time $\tau$ obeys the  relation $\cos(\xi\tau)=-1$, then no photons are found in the cavity. In addition, if we further assume that $\omega_m\tau=\pi$ then the position displacement of the oscillator produced by $\hat{U}_{OM}(\tau)$ is maximum. This implies that $\xi=(2n+1)\omega_m$, where $n$ is a large integer number, which is consistent with the assumptions underlying hamiltonian (\ref{eq:HI}). In this scenario, the normalized state of the oscillator is
\begin{eqnarray}\label{eq:MeterState}
\ket{\psi}=\frac{1}{2\sqrt{P}}\Big[\delta\ket{0}-(r/\sqrt{2})\ket{\varphi}+(t/\sqrt{2})\ket{-\varphi}\Big],
\end{eqnarray}
where $\varphi=g_0/\omega_m$  is a scaled optomechanical strength and $\ket{\pm\varphi}$ are coherent states (notice that the subscript $m$ has been dropped out). $P=\delta^2+\varphi^2/4$ is the probability of post-selection. The reflection and transmission coefficients can be written in terms of the post-selection parameter $\delta$ as  $r=(\sqrt{1-\delta^2}-\delta)/\sqrt{2}$ and $t=(\sqrt{1-\delta^2}+\delta)/\sqrt{2}$, where $\delta\in[-1/\sqrt{2},1/\sqrt{2}]$. 

From (\ref{eq:MeterState}) we see that the oscillator is in a superposition of three states that are not orthogonal since $\varphi\ll1$. In this state, the expectation value of $\hat{q}/x_0=(\hat{c}^{\dagger}+\hat{c})$ (the variable $\hat{q}$ is the position of the mirror), is given by
\begin{eqnarray}\label{eq:Position}
\frac{\langle\hat{q} \rangle}{x_0}=(2\varphi)f, \quad f=\frac{-\delta\sqrt{1-\delta^2}}{2P}.
\end{eqnarray}
Here, $f$ is an amplification factor due to the post-selection and we have adopted  the same notation as in \cite{Feizpour2011}.  Notice that the mean position of the oscillator is displaced to the right when $\delta$ is negative and to the left when it is positive. 

When $|\delta|\gg\varphi$, then $P\approx \delta^2=|\langle f|i\rangle|^2$, i.e. the back-action caused by the optomechanical interaction on the system (the photon) can be neglected. This situation corresponds to the \textit{weak measurement regime} and when this condition is not fulfilled we will say that the measurement is strong. For a weak measurement,  the amplification factor is $-\sqrt{1-\delta^2}/2\delta$, that corresponds to the weak value of $\hat{N}$ between the states $\ket{i}$ and $\ket{f}$, as can be checked by a straightforward application of the definition (\ref{eq:TheWV})
\begin{eqnarray}\label{eq:WeakValue}
N_w=\frac{\bra{f}\hat{N}\ket{i}}{\langle f|i \rangle}=-\frac{\sqrt{1-\delta^2}}{2\delta},
\end{eqnarray}
where $\ket{i}=-(1/\sqrt{2})(\hat{l}_1^{\dagger}\ket{\emptyset}_T+\hat{r}_2^{\dagger}\ket{\emptyset}_T)\ket{\emptyset}_c$. Notice that $P\sim\delta^2$ whereas the weak value $N_w$ (or the amplification factor)  $\sim\delta^{-1}$ when $\delta\ll1$.  Also, the meter state (\ref{eq:MeterState}) reduces to a coherent state $\ket{\varphi N_w}$, i.e. its original wave function is only translated, according to equation (\ref{DerivationofWV}). 

\begin{figure}
 \centering \includegraphics[width=\linewidth]{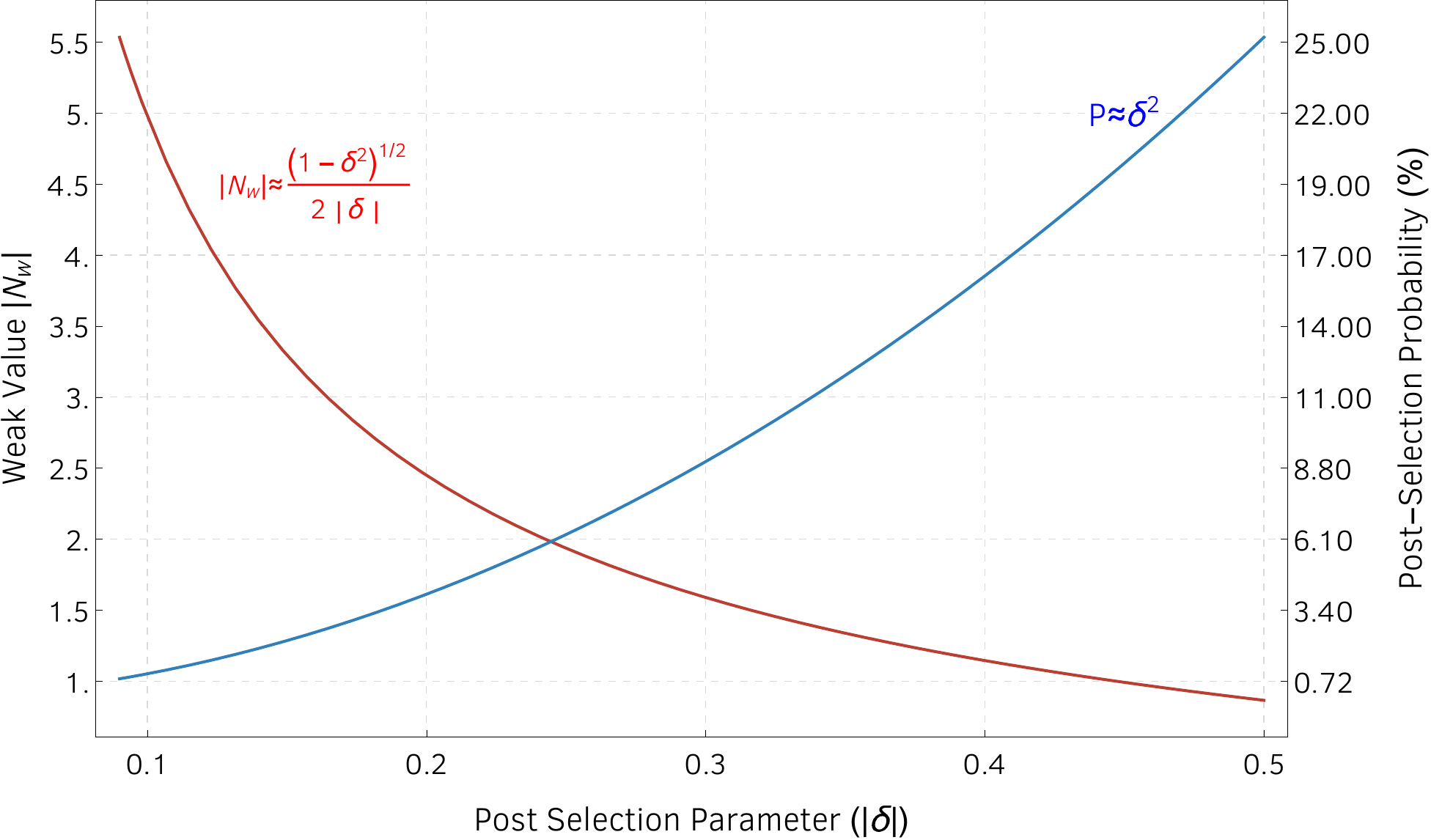}
 \caption{The magnitude of the weak value (red curve, left vertical axis) and the probability of post-selection (blue curve, right vertical axis) are plotted against the magnitude of the post-selection parameter $\delta$ for the case $\varphi=10^{-4}$.} \label{fig:fig3}
\end{figure}

In this regime, the oscillator is equally affected by the initial state $\ket{i}$ and the final state $\ket{f}$ of the photon. We observe that the pre and post-selection of the photonic state acts as if we had a scenario with many photons without post-selection.  For example, when $\varphi\sim10^{-4}$ we can set $\delta=10^{-2}$ and displace the oscillator as if the difference of photons between both sides of the cavity was of 50. Table \ref{table:table1} and figure \ref{fig:fig3} present the weak value and the post-selection probability for $|\delta|\sim10^{-1}$. In this case, the weak value of the number of photons goes from 1 to 5, with post-selection probabilities in the range of $1-25\%$.

\begin{table}
\begin{tabular}{ c | c | c }
\hline
$|\delta|$ & $|N_w|$ & $P (\%)$ \\
\hline
$0.5$ & 0.9 & 25 \\
$0.4$ & 1.1 & 16 \\
$0.3$ & 1.6 & 9 \\
$0.2$ & 2.4 & 4 \\
$0.1$ & 5.0 & 1 \\
$0.09$ & 5.5 & 0.81 \\
  \hline
\end{tabular}
\caption{As the magnitude of the dark port post-selection parameter $\delta$ decreases, the magnitude of the weak value of the radiation force $N_w$ is increased whereas the probability of post-selection decreases quadratically.  Notice that WVA (weak values outside the spectrum of $\hat{N}$) begins when $|\delta|$ is slightly below 0.5. Recall also that the weak measurement regime occurs when $\varphi\ll |\delta|$. }
\label{table:table1}
\end{table}

\begin{figure}
 \centering \includegraphics[width=\linewidth]{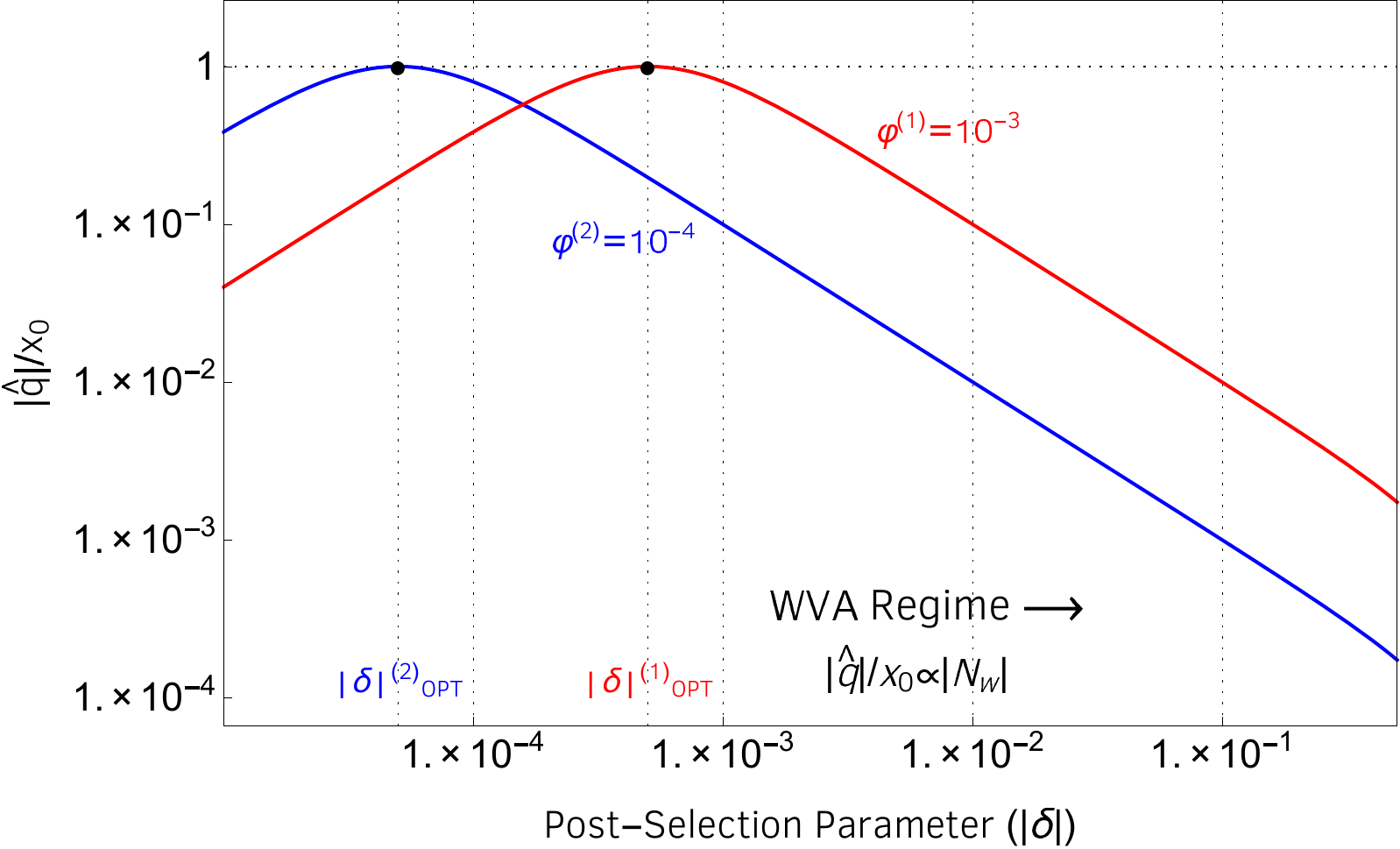}
 \caption{The average position of the oscillator divided by the zero-point fluctuation is plotted against the magnitude of the post-selection parameter. The scaled optomechanical strengths are  $\varphi^{(2)}=10^{-4}$  (blue curve) and $\varphi^{(1)}=10^{-3}$ (red curve). The oscillator reaches the level of the zero-point fluctuation \textit{outside} the weak measurement regime, when $|\delta|^{(i)}_{OPT}=\varphi^{(i)}/2, i=1,2$. Nevertheless, the probabilities of post-selection associated to this high amplification factor are extremely small, since $P\sim\varphi^2$.} \label{fig:fig4}
\end{figure}

Notice that the weak value is restricted by $|N_w|\ll\varphi^{-1}$ in order to neglect the back-action on the system. From equation (\ref{eq:Position}) it is clear that WVA does not reach the level of vacuum fluctuations. This means that lot of data is required to estimate the centroid of the probability distribution given by state ($\ref{eq:MeterState})$, i.e. the quantum-limited SNR is below unity.  In particular, the modulus of the amplification factor $|f|$ is maximized when $|\delta|=\varphi/2$, outside the weak measurement regime. For this particular post-selection parameter, the oscillator reaches the level of vacuum fluctuations, as it is shown in figure (\ref{fig:fig4}). In this case, there is a high amplification factor $\sim\varphi^{-1}$, but the probabilities of post-selection are very small, $P\sim\delta^2$. This amplification was theoretically obtained by Li \textit{et al.} \cite{Li2014} in a different interferometric setup that involves two cavities and where the amplification appears when the non linear Kerr term of the optomechanical interaction is retained. In our setup, such amplification occurs outside the regime of weak measurements and the non linear Kerr phase has no effect. 

\section{Final Remarks}\label{sec:FinalRemarks}

\begin{figure}
 \centering \includegraphics[width=\linewidth]{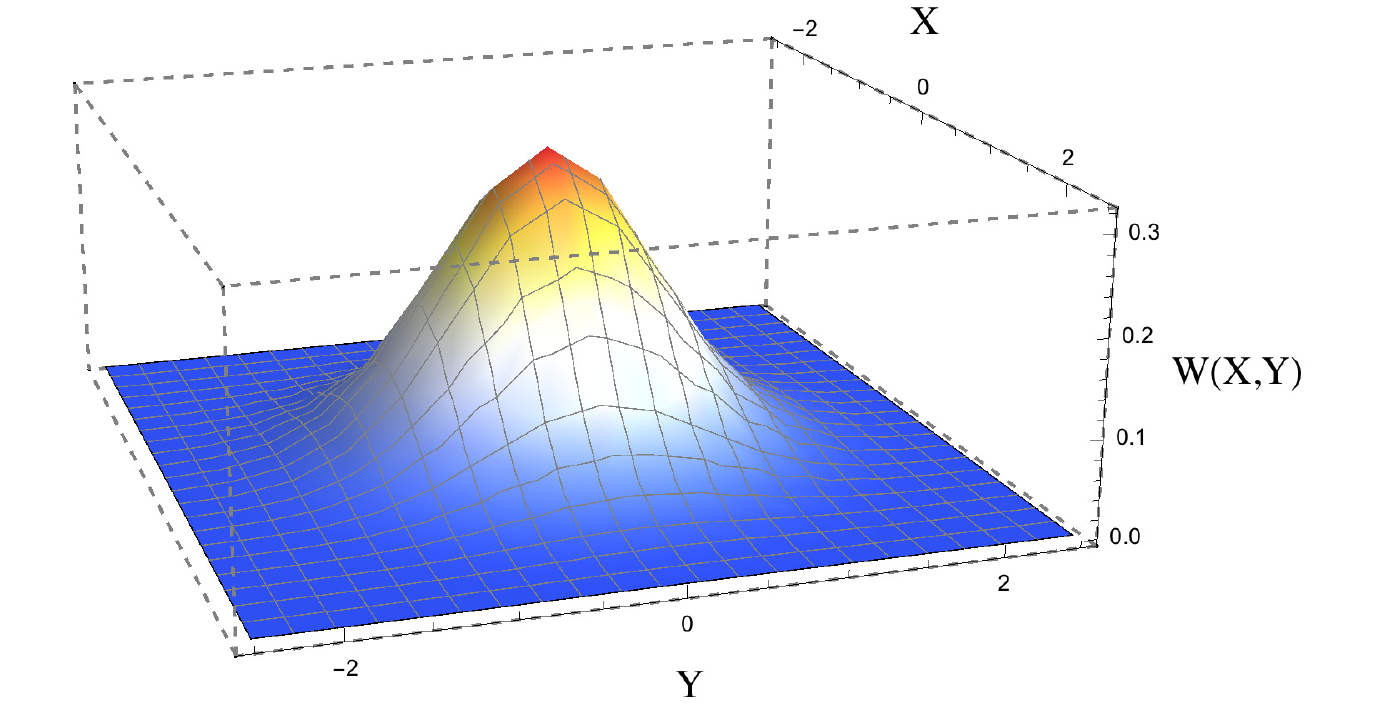}
 \caption{Wigner function of the final state of the measuring device (the oscillator) in the weak measurement regime . In this case, the scaled optomechanical strength is $\varphi=10^{-3}$ and the post-selection parameter $\delta=5\cdot10^{-2}$. Here (and in figure \ref{fig:fig6}) $\hat{X}=(\hat{c}+\hat{c}^{\dagger})/\sqrt{2}$ and $\hat{Y}=i(\hat{c}^{\dagger}-\hat{c})/\sqrt{2}$ are the mirror quadratures, satisfying $[\hat{X},\hat{Y}]=i$.} \label{fig:fig5}
\end{figure}

\begin{figure}
 \centering \includegraphics[width=\linewidth]{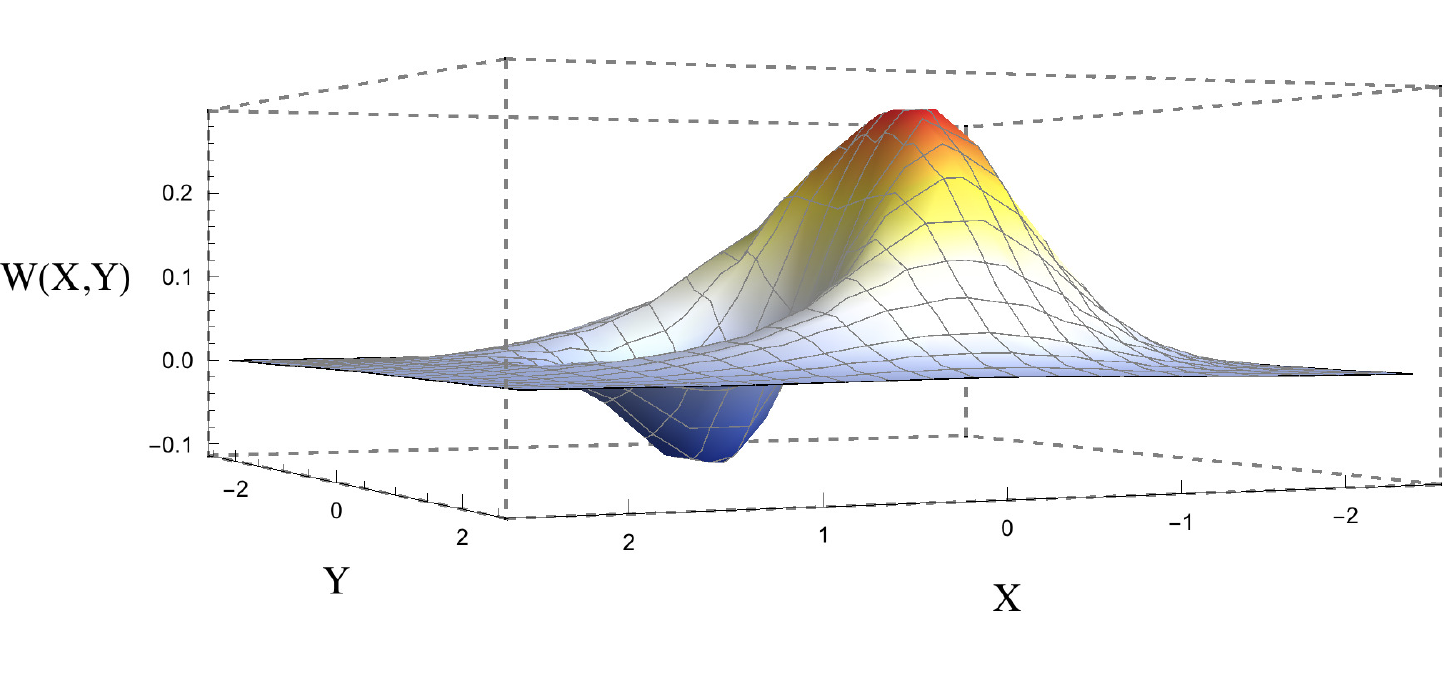}
 \caption{Wigner function of the final state of the oscillator \textit{outside} the weak measurement regime, for the case of maximum amplification and displacement to the left ($\delta=\varphi/2$).  The optomechanical strength $\varphi$ is set to be $10^{-3}$.} \label{fig:fig6}
\end{figure}

Regarding the oscillator, it is worth mentioning that this setup generates non classical mechanical states, which consist of a superposition between the ground state and the one-phonon state. Since $\varphi\ll1$, a first-order expansion of the coherent states $\ket{\pm\varphi}$ allows to express state (\ref{eq:MeterState}) as
\begin{eqnarray}\label{eq:MeterState2}
\ket{\psi}=\frac{1}{2\sqrt{P}}(2\delta\ket{0}-\varphi\sqrt{1-\delta^2}\ket{1}).
\end{eqnarray}
In the weak measurement regime, this state becomes $\ket{0}+(\varphi N_w)\ket{1}$, i.e. a ``weak'' superposition between the ground state and the first excited state. The Wigner function of this state, shown in figure (\ref{fig:fig5}), has no negative part and thus can be understood as a classical state.  As it is pointed out in \cite{Simon2011}, this slight superposition between the ground state and the one-phonon state can give rise to a large amplification effect, without changing the shape of initial wave function of the measuring device, but only translating it.  On the other hand, outside the weak measurement regime, e.g when $\delta=\varphi/2$ (maximum amplification and displacement to the left), then the state of the oscillator is  $\frac{1}{\sqrt{2}}(\ket{0}-\ket{1})$, an equal superposition of the ground state and the first excited state (see figure \ref{fig:fig6}). In this situation, the amplification factor is larger, but the wave function of the oscillator is no longer Gaussian, due to the larger weight of the one-phonon state, and the post-selection probabilities are smaller. 

In the regime of WVA, the amplification factor $f$ corresponds to the weak value $N_w\approx-1/2\delta$. The operator $\hat{N}$, defined by equation (\ref{eq:Force}), can be expressed as $\hat{N}_1-\hat{N}_2$, where $\hat{N}_i=\hat{a}^{\dagger}_i\hat{a}_i+\hat{b}^{\dagger}_i\hat{b}_i$ and $i=1,2$ (the number of interacting photons in each side). Since the weak value of a sum of operators corresponds to the sum of the weak values, then $N_{1,w}-N_{2,w}=-1/2\delta$, where $N_{i,w}$ is the weak value of $\hat{N}_i$. On the other hand, from the definition (\ref{eq:TheWV}), it is easy to see that the weak value of the total number of photons is 1, i.e. $N_{1,w}+N_{2,w}=1$. Therefore, $N_{1,w}=1/2-1/4\delta$ and $N_{2,w}=1/2+1/4\delta$. Accordingly, the amplification effect in the WVA regime shows that the weak number of photons in one side is much larger than $1$ whereas the weak number of photons in the other side is much smaller than $-1$. Both anomalous weak values contribute to the amplification of the radiation effect on the oscillator. In particular, negative weak values of number operators revert the sign of the interaction \cite{TSVF2008}, displacing the oscillator to the opposite direction that one would expect. 

In summary, we have proposed an experiment that consists of an optical cavity, with a high-Q mechanical oscillator in the middle, mounted inside an interferometer, and operated at the level of single photons. We have presented a hamiltonian model  for the case in which the injected photons are nearly monochromatic, the scaled optomechanical strength is weak ($\varphi=g_0/\omega_m\ll1$) and the system is outside the resolved sideband regime ($\omega_m/\xi\ll1$). By performing unitary evolution we have shown that, when the magnitude of the dark port post-selection parameter $\delta$ is larger than $\varphi$, the interaction between the photon and the mechanical oscillator, followed by the post-selection, is a weak measurement of the radiation force, i.e. it shifts the initial Gaussian wave function of the oscillator by a quantity proportional to the weak value of the radiation force. This produces an amplification of the mean position of the oscillator, as compared to the no post-selection scenario. When $|\delta|$ is small, the amplification is of the order of $|\delta|^{-1}$ and the post-selection probability $P\sim\delta^2$. As an example, if $\delta=-0.15$, then the weak value of the difference of photons is approximately equal to 3, and the associated probability of post-selection is nearly $3\%$.  The proposed setup might be useful for the estimation of $g_0$, when the oscillator is operated at the quantum level. 

In the weak measurement regime, we have explained that the amplification occurs because weak values of the radiation force operator $\hat{N}$ exceed the range of eigenvalues. On the other hand, we have  seen that, by choosing $\delta\sim \varphi$, the amplification factor is increased, but also the level of back-action and the measurement is no longer weak in the sense of \cite{AAV1988}, with much lower associated probabilities as compared to the weak measurement regime.  

\section*{ACKNOWLEDGMENTS}
We thank the financial support of Conicyt with the project Fondecyt $\#$1180175.

\appendix

\section{Hamiltonian Approximation}\label{app:A}

It is convenient to introduce the angular momentum operators
\begin{eqnarray}\nonumber
\hat{J}_{x1}=\frac{\hbar}{2}(\hat{a}_1\hat{b}_1^{\dagger}+\hat{a}_1^{\dagger}\hat{b}_1), \quad \hat{J}_{x2}=\frac{\hbar}{2}(\hat{a}_2\hat{b}_2^{\dagger}+\hat{a}_2^{\dagger}\hat{b}_2),\\ \nonumber
\hat{J}_{y1}=i\frac{\hbar}{2}(\hat{a}_1\hat{b}_1^{\dagger}-\hat{a}_1^{\dagger}\hat{b}_1),\quad\hat{J}_{y2}=i\frac{\hbar}{2}(\hat{a}_2^{\dagger}\hat{b}_2-\hat{a}_2\hat{b}_2^{\dagger}),\\ 
\hat{J}_{z1}=\frac{\hbar}{2}(\hat{a}_1^{\dagger}\hat{a}_1-\hat{b}_1^{\dagger}\hat{b}_1),\quad\hat{J}_{z2}=\frac{\hbar}{2}(\hat{b}_2^{\dagger}\hat{b}_2-\hat{a}^{\dagger}_2\hat{a}_2).
\end{eqnarray}
With this notation hamiltonian (\ref{eq:HI}) becomes
\begin{eqnarray}\label{eq:HI3}
\hat{H}_{I}&=&2\xi\hat{J}_{x}+\hbar\omega_{m}\hat{c}^{\dagger}\hat{c}-g_0\Big(
\frac{\hbar}{2}\hat{N}+\hat{J}_{z}\Big)(\hat{c}^{\dagger}+\hat{c}),
\end{eqnarray}
where $\hat{N}$ is defined in (\ref{eq:Force}), $\hat{J}_x=\hat{J}_{x1}+\hat{J}_{x2}$ and $\hat{J}_z=\hat{J}_{z1}+\hat{J}_{z2}$. Notice that the radiation pressure interaction can be further split into two terms, i.e. $\hat{H}_{\rm{OM}1}=-\frac{g_0\hbar}{2}\hat{N}(\hat{c}^{\dagger}+\hat{c})$ and $\hat{H}_{\rm{OM}2}=-g_0\hat{J}_{z}(\hat{c}^{\dagger}+\hat{c})$. In order to see the time dependence of the second process we go to a rotating frame with respect to $2\xi\hat{J}_{x}+\hbar\omega_{m}\hat{c}^{\dagger}\hat{c}+\hat{H}_{\rm{OM}1}$. In this picture, the hamiltonian is given by
\begin{eqnarray}\nonumber
\hat{H}_{II}(\tau)=
-\hat{J}_z[\hat{c}^{\dagger}A(\tau)+\hat{c}A^*(\tau)+\hat{N}f(\tau)]\\
-\hat{J}_{y}[\hat{c}^{\dagger}B(\tau)+\hat{c}B^*(\tau)+\hat{N}g(\tau)],
\end{eqnarray}
where the time dependent coefficients, with units of frequency, are
\begin{eqnarray}\nonumber
A(\tau)=g_0\cos(2\xi \tau)e^{i\omega_m \tau},\text{ } f(\tau)=\frac{g_0^2}{\omega_m}\cos(2\xi \tau)[1-\cos(\omega_m\tau)],\\ \nonumber
 B(\tau)=g_0\sin(2\xi \tau)e^{i\omega_m \tau},\text{ }  g(\tau)=\frac{g_0^2}{\omega_m}\sin(2\xi \tau)[1-\cos(\omega_m\tau)], \\
\end{eqnarray}
and $A^*$ and $B^*$ denote complex conjugation. The time evolution operator in this frame is given by the \textit{Dyson} series, that is,
\begin{eqnarray}\nonumber
\hat{U}_{II}(\tau)=1-\frac{i}{\hbar}\int_0^\tau\rm{d} t_1\hat{H}_{II}(t_1)+\\
\Big(\frac{-i}{\hbar}\Big)^2\int_{0}^\tau\rm{d} t_2\hat{H}_{II}(t_2)\int_0^{t_2}\rm{d} t_1\hat{H}_{II}(t_1)+...
\end{eqnarray}
The first order term is given by
 \begin{eqnarray}\label{1stDyson}\nonumber
\hat{U}_{II}^{(1)}(\tau)= i(\hat{J}_{z}/\hbar)\Big[\hat{c}^{\dagger}\bar{A}(\tau)+\hat{c}\bar{A}^*(\tau)+\hat{N}\bar{f}(\tau)\Big]+\\
 i(\hat{J}_{y}/\hbar)\Big[\hat{c}^{\dagger}\bar{B}(\tau)+\hat{c}\bar{B}^*(\tau)+\hat{N}\bar{g}(\tau)\Big].
 \end{eqnarray}
The overline indicates integration in time of the coefficients, namely, 
\begin{eqnarray}\nonumber
&&\bar{A}(\tau)=\int_{0}^{\tau}A(z)\rm{d} z\\ \nonumber
&&= -i\Big(\frac{g_0}{2\xi}\Big)\Big(  \frac{\omega_{m}}{2\xi}    \Big)
\Big[ \frac{1}{1-(\omega_{m}/2\xi)^2}\Big]\Big[1-\cos(2\xi \tau)e^{i\omega_{m}\tau}\Big]\\ 
&&+\Big(\frac{g_0}{2\xi}\Big)
\Big[ \frac{1}{1-(\omega_{m}/2\xi)^2}\Big]\sin(2\xi \tau)e^{i\omega_{m}\tau},
 \end{eqnarray}
\begin{eqnarray}\nonumber
&&\bar{B}(\tau)=\int_{0}^{\tau}B(z)\rm{d} z\\ \nonumber
&&=\Big(\frac{g_0}{2\xi}\Big)
\Big[ \frac{1}{1-(\omega_{m}/2\xi)^2}\Big]\Big[1-\cos(2\xi \tau)e^{i\omega_{m}\tau}\Big]\\
&&+i\Big(\frac{g_0}{2\xi}\Big)\Big( \frac{\omega_{m}}{2\xi}    \Big)
\Big[ \frac{1}{1-(\omega_{m}/2\xi)^2}\Big]\sin(2\xi \tau)e^{i\omega_{m}\tau},
 \end{eqnarray}
\begin{eqnarray}\nonumber
&&\bar{f}(\tau)=\int_{0}^{\tau}f(z)\rm{d} z\\ \nonumber
&&=\Big(\frac{g_0}{\omega_{m}}\Big)
\Big(\frac{g_0}{2\xi}\Big)\sin(2\xi \tau)\\ \nonumber
&&-\Big(\frac{g_0}{2\xi}\Big)
\Big(\frac{g_0}{\omega_{m}}\Big)
\Big[ \frac{1}{1-(\omega_{m}/2\xi)^2}\Big]\cos(\omega_{m}\tau)\sin(2\xi \tau)\\
&&+\Big(\frac{g_0}{2\xi}\Big)^2\Big[ \frac{1}{1-(\omega_{m}/2\xi)^2}\Big]\cos(2\xi \tau)\sin(\omega_{m}\tau),
\end{eqnarray}
and
\begin{eqnarray}\nonumber
&&\bar{g}(\tau)=\int_{0}^{\tau}g(z)\rm{d} z \\ \nonumber
&&=\Big(\frac{g_0}{\omega_{m}}\Big)
\Big(\frac{g_0}{\xi}\Big)\sin^2(2\xi \tau)\\ \nonumber 
&&-\Big(\frac{g_0}{2\xi}\Big)
\Big(\frac{g_0}{\omega_{m}}\Big)
\Big[ \frac{1}{1-(\omega_{m}/2\xi)^2}\Big]\\ \nonumber
&&+\Big(\frac{g_0}{\omega_{m}}\Big)
\Big(\frac{g_0}{2\xi}\Big)
\Big[ \frac{1}{1-(\omega_{m}/2\xi)^2}\Big]\cos(\omega_{m}\tau)\cos(2\xi \tau)\\
&&+\Big(\frac{g_0}{2\xi}\Big)^2\Big[ \frac{1}{1-(\omega_{m}/2\xi)^2}\Big]\sin(2\xi \tau)\sin(\omega_{m}\tau).
\end{eqnarray}
There are three expansion parameters appearing in the terms above, which are $g_0/\xi$, $g_0/\omega_{m}$ and $\omega_{m}/\xi$. We will assume that all these terms $\ll1$, which can be summarized by the condition $g_0\ll\omega_m\ll \xi$, i.e. the parameter $g_0/\xi$ is the smallest term. We will neglect terms $\sim g_0/\xi$, which entails that $\bar{A}(\tau)\approx\bar{B}(\tau)\approx\bar{f}(\tau)\approx\bar{g}(\tau)\approx0$ and accordingly all the remaining terms of the series are also negligible. Thus, in this regime the evolution operator is just the identity. Since kets in this frame do not change, the evolution given by $\hat{H}_{\rm{OM}2}$ can be disregarded from the hamiltonian in the previous picture, and the hamiltonian reduces to (\ref{eq:HI2}).

\section{Time Evolution Operator}\label{app:B}

Since $\hat{J}_x$ commutes with the other terms in hamiltonian (\ref{eq:HI2}), the corresponding evolution operator is
\begin{eqnarray}\nonumber
\hat{U}_I(\tau)&=&\exp{[-i\omega_m\tau\hat{c}^{\dagger}\hat{c}+i(g_0/2)\tau\hat{N}(\hat{c}^{\dagger}+\hat{c})]}\\ \label{eq:TimeEvolutionOperator2}
&&\times\exp{[-i(2\xi \tau/\hbar)\hat{J}_x]}
\end{eqnarray}
The first term can be worked out exactly in the same way as in \cite{Bose1997}, which produces the final form presented in equation (\ref{eq:TimeEvolutionOperator}).


\end{document}